\documentclass[a4paper,fleqn,usenatbib]{mnras}
\usepackage{newtxtext,newtxmath}

\usepackage[T1]{fontenc}
\usepackage{ae,aecompl}

\usepackage{bm}

\usepackage{graphicx}	
\usepackage{amsmath}	
\usepackage{amssymb}	

\title[Self-trapping of g-mode oscillations]{Self-Trapping of G-Mode Oscillations in Relativistic Thin Disks, Revisited II: Revision of Boundary Condition}

\author[Shoji Kato]{
Shoji Kato $^{1}$\thanks{E-mail: kato.shoji@gmail.com}
\\
$^{1}$2-2-2 Shikanodai Nishi, Ikomashi, Nara, Japan, 630-0114}

\date{Accepted XXX. Received YYY; in original form ZZZ}

\pubyear{2017}

\begin{document}
\label{firstpage}
\pagerange{\pageref{firstpage}--\pageref{lastpage}}
\maketitle

\begin{abstract}In a previous paper (Kato 2017a) we have examined how the self-trapping of g-mode oscillations in 
geometrically thin relativistic disks is affected by the presence of uniform vertical magnetic fields.
Disks considered are isothermal in the vertical direction and are truncated at a
certain height by presence of hot corona.
After a correction of simple analytical error, we showed (Kato 2017b) that the self-trapping 
of axisymmetric g-mode oscillations in non-magnetized disks is destroyed by weak magnetic fields as Fu and Lai (2009) showed. 
In this paper, however, we re-examine the same problem by imposing a different more
relevant boundary condition on the disk-corona surface and find that the 
self-trapping of axisymmetric g-mode oscillations seems to still exist like the case of 
non-magnetized disks.
\end{abstract}

\begin{keywords}
accretion, accretion discs -- black hole physics -- magnetic fields -- waves -- X-ray binaries 
\end{keywords}



\section{Introduction}

In black-hole and  neutron-star X-ray binaries quasi-periodic oscillations (QPOs) are occasionally 
observed with frequencies close to the relativistic Keplerian frequencies of the innermost region of surrounding disks 
(e.g., van der Klis 2000; Remillard \& McClintock 2006). 
One of possible origins of these quasi-periodic oscillations is disk oscillations (discoseismology)(e.g., Wagoner 1999, Kato 2001, Kato 2016).

The innermost regions of relativistic disks are subject to strong gravitational field of 
central sources, and thus the radial distribution of (radial) epicyclic frequency 
is quite different from that 
in the Newtonian disks (Aliev \& Galtsov 1981).
Hence, disk oscillations in these relativistic regions are quite
different from those in Newtonian disks (Kato and Fukue 1980).
In particular,  Okazaki et al. (1987) showed that the g-mode oscillations\footnote
{See, for example, Kato (2016) for classification of disk oscillation modes.
}
whose frequency $\omega$ is lower than the 
maximum value of epicyclic frequency,  say $\kappa_{\rm max}$, are self-trapped in the finite region bounded between two radii 
where $\omega=\kappa$ are realized.
This is the self-trapping of the g-mode oscillations in relativistic disks.
Eigen-frequencies of  these trapped g-mode oscillations in geometrically thin disks are calculated in detail 
by Perez et al. (1997), using  the Kerr geometry.
Nowak et al. (1997) suggested that these trapped oscillations might be the origin of the 67 Hz oscillations observed in the black-hole source GRS 1915+105.

In the above studies, however, effects of magnetic fields on oscillations were  outside considerations.
Fu and Lai (2009) examined these effects and found important results.
That is, they showed that the g-mode oscillations are strongly affected if poloidal magnetic fields
are present in disks and the self-trapping of g-mode oscillations is destroyed even when the fields are weak. 
Their analyses, however, are rough in examining effects of vertical disk structure 
on g-mode oscillations by the following reasons. 
In geometrically thin disks with poloidal magnetic fields, the disks are  strongly inhomogeneous in the vertical direction.
That is, density sharply decreases from the equatorial plane in the vertical direction, 
while the Alfv\'{e}n speed increases greatly in the direction.
These strong inhomogeneous structures in the vertical direction should be considered 
in examining the behavior of g-mode oscillations, since g-mode oscillations are those which have node(s) in the vertical direction.  
Recently, Ortega-Rodriguez et al. (2015) reexamined the effects of magnetic fields
on trapping of the g-mode oscillations.
Their analyses, however, are still in the framework of
the WKB approximations. 

To avoid mathematical complication and difficulty related to the above strong inhomogeneity in the vertical direction,
Kato (2017a, referred to paper I) considered disks with a finite vertical thickness by 
truncation due to a hot corona.
This adoption of truncated disks is, however, not only for avoidance of 
mathematical complication, but also for physical considerations. 
That is, many black-hole and neutron-star X-ray binaries show both soft and hard
components  of spectra, and record a soft-hard transition in their X-ray spectra.
Usually three states are known:
high/soft state, very high/intermediate state (steep power-law state), and low/hard state.
Remillard (2005) shows that detection of high frequency QPOs in 
black-hole X-ray binaries is correlated with the power-law luminosity, and their frequencies are related to the soft component.
This suggests that at the state where high frequency QPOs are
observed, the sources have both geometrically thin disks and hot coronae.

The analyses in paper I, however, seem to be unsatisfactory.
That is, the boundary condition adopted at the disk-corona transition layer 
is not always proper.
Supplementary, the zeroth order eigenfunctions used to expand perturbed parts of
eigenfunctions are not rigorously orthogonal.
In this context, in this paper we re-examine the issue of the trapping of the g-mode 
oscillations with a different boundary condition which will be more proper than that
used in paper I.
The results derived by using this revised boundary condition show the presence of
self-trapping of g-mode oscillations, unlike those in paper I.   
This suggests that the self-trapping of g-mode 
oscillations is still one of possible candidates of QPOs.
In addition, the fact that the trapping of oscillations is rather 
sensitive to boundary condition suggests importance of further 
studies on what boundary conditions are realistic,  
in order to finally judge whether
the trapping of g-mode oscillations is really present or not.

Structure of this paper is almost the same as that in paper 1.
Arguments on boundary condition between disk and corona are given in section 2.3.
Orthogonality of zeroth order eigenfunctions is presented in section 3.1.
The wave equation derived by taking the effects of magnetic fields into account as 
perturbations are given in section 4.
The trapping regions of oscillations in propagation diagram are presented in 
section 5 for various cases of disk parameters and oscillation modes.

\section{Unperturbed Disk Model and Basic Equations Describing Perturbations}

Unperturbed disk model and basic equations describing perturbations are presented here for completeness, although
they are the same as those in Paper I.
First we mention that, following a conventional way, 
we treat Newtonian equations, but adopt the
general relativistic expression for (radial) epicyclic frequency $\kappa(r)$.
 
Disks are assumed to be vertically isothermal and  geometrically thin, subjected to uniform vertical magnetic fields.
We adopt the cylindrical coordinates ($r$, $\varphi$, $z$) whose origin is at the center of a central object 
and the $z$-axis is perpendicular to the disk plane.
Then, the magnetic fields in the unperturbed state, $\bm{B}_0$, are  
\begin{equation}
       \bm{B}_0(r)=[0, 0, B_0(r)].
\label{unifom-fields}
\end{equation}

Since the magnetic fields are purely vertical, they have no effects on the vertical structure of disks,
and the hydrostatic balance in the vertical direction in disks gives the disk density, $\rho_0$, stratified in the vertical 
direction as (e.g., see Kato et al. 2008)
\begin{equation}
          \rho_0(r, z)=\rho_{00}(r)\, {\rm exp}\biggr(-\frac{z^2}{2H^2}\biggr),
\label{density}
\end{equation}
where the scale height, $H(r)$, is related to the isothermal acoustic speed, $c_{\rm s}(r)$, and
vertical epicyclic frequency, $\Omega_\bot(r)$, by
\begin{equation}
       H^2(r)=\frac{c_{\rm s}^2}{\Omega_\bot^2}.
\label{scaleheight}
\end{equation}
In relativistic disks under the Schwarzschild metric, $\Omega_\bot$ is equal to the
relativistic Kepler frequency.

\subsection{Truncation of disk thickness by hot corona}

The above-mentioned vertically isothermal disks can extend infinitely in the vertical direction, 
although they are derived under the assumption that the disks are thin in the vertical direction.
Here, we assume that the disks are terminated at a finite height, say $z_{\rm s}$, by presence of hot corona.
The height $z_{\rm s}$ is taken to be a parameter.
As a dimensionless parameter we adopt $\eta_{\rm s}\equiv z_{\rm s}/H$ hereafter.

\subsection{Equations describing perturbations}

We consider axisymmetric, small-amplitude adiabatic perturbations in the above-mentioned vertically isothermal disks.
The perturbations are assumed to be isothermal in the vertical direction.
Furthermore, the perturbations are time-periodic with frequency $\omega$, i.e., $\partial/\partial t=i\omega$.

Eulerian velocity perturbations over the rotation, $r\Omega(r)$, are denoted by ($u_r$, $u_\varphi$, $u_z$), and
perturbations of magnetic fields over the uniform vertical fields, $\bm {B}_0$, are denoted by 
$(b_r, b_\varphi, b_z)$.
Then, the $r$-, $\varphi$-, and $z$-components of equations of motion describing the perturbations are, respectively,
\begin{equation}
   i\omega u_r-2\Omega u_\varphi=-\frac{\partial h_1}{\partial r}+\frac{c_{\rm A}^2}{B_0}
                          \biggr(\frac{\partial b_r}{\partial z}-\frac{\partial b_z}{\partial r}\biggr),
\label{ur}
\end{equation}
\begin{equation}
      i\omega u_\varphi+\frac{\kappa^2}{2\Omega}u_r=\frac{c_{\rm A}^2}{B_0}\frac{\partial b_\varphi}{\partial z},
\label{uvarphi}
\end{equation}
\begin{equation}
      i\omega u_z=-\frac{\partial h_1}{\partial z},
\label{uz}
\end{equation}
where $h_1$ is related to pressure and density variations, $p_1$ and $\rho_1$, by
\begin{equation}
           h_1=\frac{p_1}{\rho_0}=c_{\rm s}^2\frac{\rho_1}{\rho_0},
\label{h1}
\end{equation}
and $c_{\rm A}$ is the Alfv\'{e}n speed defined by
\begin{equation}
       c_{\rm A}^2(r, z)=\frac{B_0^2}{4\pi\rho_0}=c_{{\rm A}0}^2(r)\,{\rm exp}\biggr(\frac{z^2}{2H^2}\biggr),
\label{Alfven}
\end{equation}
$c_{{\rm A}0}$ being the Alfv\'{e}n speed on the equator. 

The time variation of magnetic fields is governed by the induction equation, whose $r$-, $\varphi$-, and $z$-components
are, respectively,
\begin{equation}
      i\omega b_r=B_0\frac{\partial u_r}{\partial z},
\label{br}
\end{equation}
\begin{equation}
      i\omega b_\varphi=B_0\frac{\partial u_\varphi}{\partial z}+r\frac{d\Omega}{dr}b_r,
\label{bvarphi}
\end{equation}
\begin{equation}
      i\omega b_z=-B_0\frac{\partial}{r\partial r}(ru_r).
\label{bz}
\end{equation}

Finally, the time variation of density is governed by the equation of continuity, which is
\begin{equation}
     i\omega \rho_1
        +\frac{\partial}{r\partial r}(r\rho_0u_r)+\frac{\partial}{\partial z}(\rho_0u_z)=0.
\label{continuity}
\end{equation}

Hereafter, radial variations of unperturbed quantities, 
such as $\Omega(r)$, $\kappa(r)$,
$\rho_{00}(r)$, $H(r)$, $c_{{\rm s}}^2(r)$
and $c_{{\rm A}0}^2(r)$, are neglected, assuming that radial wavelengths of 
perturbations are shorter
than the characteristic radial lengths of unperturbed quantities (local approximations
in the radial direction).

Our purpose here is to derive a wave equation expressed in terms of $h_1$.
To do so, we first eliminate $b_r$, $b_\varphi$, and $b_z$ from the equation of motions by using induction equations
(\ref{br}), (\ref{bvarphi}) and (\ref{bz}).
From equations (\ref{ur}) and (\ref{uvarphi}) we have, respectively,
\begin{equation}
     i\omega\, (i\omega\, u_r-2\Omega u_\varphi)=-i\omega\frac{\partial h_1}{\partial r}
            +c_{\rm A}^2\biggr(\frac{\partial^2u_r}{\partial z^2}+\frac{\partial^2u_r}{\partial r^2}\biggr),
\label{ur2}
\end{equation}
and
\begin{equation}
      -\omega^2\biggr(i\omega\, u_\varphi+\frac{\kappa^2}{2\Omega}u_r\biggr)
             =c_{\rm A}^2\biggr(i\omega\frac{\partial^2u_\varphi}{\partial z^2}
             +r\frac{d\Omega}{dr}\frac{\partial^2u_r}{\partial  z^2}\biggr).
\label{uvarphi2}
\end{equation}
Substituting $u_\varphi$ derived from equation (\ref{ur2}) into equation (\ref{uvarphi2}), we have a relation between
$h_1$ and $u_r$.
After some manipulations we rearrange the relation as
\begin{eqnarray}
&&    i\omega\biggr(\omega^2+c_{\rm A}^2\frac{\partial^2}{\partial z^2}\biggr) 
      \frac{\partial h_1}{\partial r}
              \nonumber \\
&&          =\omega^2(\omega^2-\kappa^2)u_r+c_{\rm A}^2(\omega^2-
      \kappa^2+4\Omega^2)\frac{\partial^2u_r}{\partial z^2}
                 \nonumber    \\
&&          +c_{\rm A}^2\frac{\partial^2}{\partial z^2}\biggr[c_{\rm A}
        ^2\biggr(\frac{\partial^2}{\partial r^2}+\frac{\partial^2}{\partial z^2}
              \biggr)u_r\biggr].
\label{h1ur1}
\end{eqnarray}

Another relation between $h_1$ and $u_r$ is obtained from equation of continuity (\ref{continuity}) by
using $\rho_1=\rho_0h_1/c_{\rm s}^2$ and expressing $u_z$ in terms of $h_1$ by using equation (\ref{uz}).
The result is
\begin{equation}
       i\omega\frac{\partial u_r}{\partial r}=\biggr(\frac{\partial^2}{\partial z^2}-\frac{z}{H^2}\frac{\partial}{\partial z}
                 +\frac{\omega^2}{c_{\rm s}^2}\biggr)h_1.
\label{h1ur2}
\end{equation}

We now operate with $\partial/\partial r$ on equation (\ref{h1ur1}).
Since the wavy perturbations have been assumed to have short wavelength in the radial direction (local
approximation  in the radial direction), the operator $\partial/\partial r$ acts only on $h_1$ and $u_r$.
When it is operated on $u_r$, the result, i.e., $\partial u_r/\partial r$, can be expressed in terms of $h_1$ by using equation (\ref{h1ur2}).
Then, we have an wave equation expressed in terms of $h_1$ alone.
After some manipulations the resulting equation can be expressed as
\begin{eqnarray}
&&    {\cal L}h_1+\frac{\omega^2H^2}{\omega^2-\kappa^2}\frac{\partial^2h_1}{\partial r^2}
         \nonumber \\
&&  +\frac{c_{\rm A}^2}{c_{\rm s}^2}\frac{\Omega_\bot^2}{\omega^2-\kappa^2}
   \biggr[\biggr(H^2\frac{\partial^2}{\partial r^2}+\frac{2\omega^2-(\kappa^2-4\Omega^2)}{\omega^2}
              \frac{\partial^2}{\partial\eta^2}\biggr){\cal L}h_1
              \nonumber \\
&&  \hspace{80pt} +H^2\frac{\partial^4}{\partial r^2\partial \eta^2}h_1\biggr]
         \nonumber \\
 &&+\frac{c_{\rm A}^2}{c_{\rm s}^2}\frac{\Omega_\bot^2}{\omega^2-\kappa^2}\frac{\Omega_\bot^2}{\omega^2}
       \frac{\partial^2}{\partial\eta^2}
   \biggr[\frac{c_{\rm A}^2}{c_{\rm s}^2}\biggr(H^2\frac{\partial^2}{\partial r^2}+\frac{\partial^2}{\partial\eta^2}\biggr){\cal L}h_1\biggr]
     =0,
           \nonumber \\
\label{eq-of-h1}
\end{eqnarray}
where the coordinates ($r$, $z$) are changed to ($r$, $\eta$) with $\eta\equiv z/H$, and
the operator ${\cal L}$ is defined by
\begin{equation}
      {\cal L}h_1=\biggr(\frac{\partial^2}{\partial\eta^2}-\eta\frac{\partial}{\partial\eta}+\frac{\omega^2}{\Omega_\bot^2}\biggr)h_1.
\label{calL}
\end{equation}
 
\subsection{Boundary condition}                                         
   
As mentioned before we are interested in disks which are terminated at certain height by presence 
of hot corona.
The height of termination is taken to be $z=z_{\rm s}$ (i.e., $\eta=\eta_{\rm s}
\equiv z_{\rm s}/H$).
This dimensionless disk thickness, $\eta_{\rm s}$,  is a parameter, and we
impose a boundary condition at  $\eta_{\rm s}$.

Let us consider the boundary condition more carefully than in paper I.
We start from the state where the interface has been deformed by perturbations.
The equation of motion is now integrated in a narrow width in the normal direction 
of the deformed surface. 
In the limit where the integration width is infinitely thin,
we have a fitting condition at the deformed surface, which is the continuation of  
\begin{equation}
      \rho v_n^2+p+ \frac{1}{8\rm \pi}B_t^2,
\label{BC0}
\end{equation}  
where $\rho$ and $p$ are (total) gas density and pressure, respectively, and
$v_n$ the normal component of the total velocity 
including rotation and
$B_t$ the tangential component of the total magnetic fields
(see footnote 2 of paper I).
In deriving the above equation, continuities of the normal components of $\rho {\bm v}$
and of ${\bm B}$ have been used. 

Since we are considering axially symmetric perturbations in disks where the unperturbed
magnetic fields are only in the vertical direction,
equation (\ref{BC0}) shows that $p$ is continuous at the deformed disk-corona surface,
neglecting the second order small quantities with respect to perturbations.
Before the perturbations are superposed, the pressure is continuous at the unperturbed 
disk-corona interface.
This leads to the boundary condition that the Lagrangian pressure variation, $\delta p$,
is continuous at the disk-corona interface, i.e.,  
\begin{equation}
        [[ \delta p]]=0 \quad {\rm at}\ \ \eta=\eta_{\rm s},
\label{BC-minus1}
\end{equation}
where $[[A]]$ denotes the difference of $A$ in the disk-corona interface.

Since the Lagrangian pressure variation, $\delta p$, is related to the Lagrangian density variation, $\delta\rho$, 
by $\delta p=p_0(\delta\rho/\rho_0)$, and the unperturbed pressure is continuous at the boundary, we can rewrite the boundary condition $[[\delta p]]=0$ as
\begin{equation}
         \biggr[\biggr[ \frac{\partial u_r}{\partial r}+\frac{\partial u_z}{\partial z}\biggr]\biggr]=0   \quad {\rm at}\ \ \eta=\eta_{\rm s}.
\label{BC1}
\end{equation} 
In paper I, assuming that the Lagrangian pressure variations in corona are small,
we have imposed  
\begin{equation}
       \frac{\partial u_r}{\partial r}+\frac{\partial u_z}{\partial z}=0$  \quad    {\rm at} \quad $\eta=\eta_{\rm s}
\label{BC2}
\end{equation}
to the disk gases.

The above argument on $\delta p=0$ in corona, however, seems to be not always relevant.
The argument is based on the following considerations.
For simplicity, let us consider the case where cool (dense) homogeneous gases  
and hot (thin) homogeneous gases are connected at a sharp transition layer 
with pressure balance.
There are no rotation and no magnetic fields.
In such situations we consider acoustic oscillations propagating through the 
transition layer.
Since the pressure restoring force is strong in hot region due to high temperature,
the wavelength of the oscillations in the hot region is longer than that in the cool region.
That is, ${\rm div}\, \bm{v}$ in the hot region is smaller than that in the cool region.
In the limit of large temperature difference, the hot region behaves like incompressible gases and we can thus impose 
${\rm div}\, \bm{v}=0$ as a boundary condition for cool gases at the transition radius.
 
In the present disk-corona transition problem,
the main restoring force of oscillations in (cool) disks is not the pressure one, but
that due to rotation.
Hence, wavelengths of oscillations in disks are not much shorter than those in
corona, although the temperature is low.
In other words, in principle, the amount of ${\rm div}\, \bm{v}$ is comparable 
in corona and disk.
Hence, imposing ${\rm div}\, \bm{v}=0$ to disk gases as boundary condition at 
$\eta_{\rm s}$ will not be relevant.

In this context, we return to the boundary condition (\ref{BC1}) and
derive a boundary condition which might be better than (\ref{BC2}).
If $\partial u_r/\partial r$ is not continuous at the boundary, there is a 
discontinuous shear flow on the boundary.
This will be unrealistic, and thus we impose that $\partial u_r/\partial r$ is 
continuous on the boundary.
Then, the boundary condition (\ref{BC1}) is reduced to
\begin{equation}
       \biggr[\biggr[\frac{\partial u_z}{\partial z}\biggr]\biggr]=0.
\label{BC3}
\end{equation}
In corona the vertical scale length characterizing the vertical corona structure
(i.e., the scale length corresponding to $H$ in disks) is much longer than $H$,
because of high temperature in corona.
This suggests that $\partial u_z/\partial z$ in corona is much smaller than that in
disks.
Considering this situation, we adopt $\partial u_z/\partial z=0$
as the boundary condition at  $\eta_{\rm s}$.
By using the $z$-component of the equation of motion, we can rewrite this boundary condition in terms of $h_1$ as 
\begin{equation}
          \frac{\partial^2 h_1}{\partial z^2}=0  \quad {\rm at} \quad \eta=\eta_{\rm s}.
\label{BC4}
\end{equation}
In the followings we use equation (\ref{BC4}) as the boundary condition at 
$\eta_{\rm s}$ for dis oscillations. 

It is noticed that boundary condition (\ref{BC4}) 
allows us to rigorously apply the standard perturbation method in calculations of
trapping.
First, the boundary condition (\ref{BC4}) does not involve explicitly $r$-dependent terms.
This is compatible with the method of separation of variables which will be adopted below.
Second, the zeroth-order eigenfunctions are exactly orthogonal, and thus expansion 
of the higher order perturbed quantities
by the set of the zeroth-order functions will be relevant.

\section{Perturbation Method}  

Here, we solve equation (\ref{eq-of-h1}) by a perturbation method with boundary condition (\ref{BC4}).
We start from the limit of no magnetic fields, and examine the effects of magnetic fields 
by a perturbation method.
Equation (\ref{eq-of-h1}) has terms of $c_{\rm A}^2/c_{\rm s}^2$ and $(c_{\rm A}^2/c_{\rm s}^2)^2$.
The former ones can be treated as small order perturbing quantities when 
$(c_{{\rm A}0}^2/c_{\rm s}^2)\, {\rm exp}(\eta_{\rm s}^2/2)<1$.
The latter terms are further smaller than the former ones by the same factor of 
$(c_{{\rm A}0}^2/c_{\rm s}^2)\, {\rm exp}(\eta_{\rm s}^2/2)$,
and should be neglected when a perturbation method is performed in the lowest order of approximations, as
did in paper I.

\subsection{Zeroth-order solutions and their orthogonality}

In the limit of no magnetic fields, equation (\ref{eq-of-h1}) is reduced to 
\begin{equation}
        {\cal L}h_1+\frac{\omega^2H^2}{\omega^2-\kappa^2}\frac{\partial^2h_1}{\partial r^2}=0.
\label{zerotheq}
\end{equation}
As is known, this equation can be solved by separating $h_1(r,\eta$) as $h_1(r,\eta)=g(\eta)f(r)$.
By this separation equation (\ref{zerotheq}) is  separated into two equations:
\begin{equation}
     \biggr(\frac{d^2}{d\eta^2}-\eta\frac{d}{d\eta}+K \biggr)g(\eta) =0,
\label{eqg}
\end{equation}
and 
\begin{equation}
      \frac{\omega^2H^2}{\omega^2-\kappa^2}\frac{\partial^2 f}{\partial r^2}
      +\biggr(\frac{\omega^2}{\Omega_\bot^2}-K\biggr)f=0,
\label{eqf}
\end{equation}
where $K$ is the separation constant and is determined by solving equation (\ref{eqg}) with the boundary condition;
\begin{equation}
         \frac{d^2g(\eta)}{d\eta^2}=0 \quad {\rm at} \quad \eta=\eta_{\rm s}.
\label{BC5}
\end{equation}
 
The solutions of equation (\ref{eqg}) consist of a set of eigenfunctions, $g^{(0)}_n(\eta)$, with a set of
eigenvalues, $K^{(0)}_n$, where $n$ is a positive integer, i.e., $n=1,2,...$, representing the order of 
solutions and
the superscript $(0)$ is added here in order to emphasize that the solutions presented here are
the zeroth order solutions with $c_{{\rm A}0}^2/c_{\rm s}^2=0$.

The fundamental eigenfunction of equation (\ref{eqg}), i.e., $g_1^{(0)}(\eta)$, is easily found to be
\begin{equation}
       g_1^{(0)}(\eta)=\eta \quad {\rm with} \quad K_1^{(0)}=1,
\label{g1-sol}
\end{equation}
independent of $\eta_{\rm s}$. 
This solution is the same as the fundamental eigenfunction in the case where the disk extends infinity, i.e.,
$\eta_{\rm s}=\infty$.
For $n\not= 1$, however,  the eigenfunctions in the present truncated disks are different from those of $\eta_{\rm s}=\infty$. 
The eigenvalues corresponding to $n=1,2,3,4...$ are shown in figure 1 as functions of 
$\eta_{\rm s}$.
The corresponding eigenfunctions are shown in figures 2 and 3, where
figure 2 is  for odd modes of $n=1$, $n=3$ and $n=5$, and figure 3 is for even modes 
of $n=2$ and $n=4$.
Both figures are written for two cases where $\eta_{\rm s}=3$ and $\eta_{\rm s}=5$.
 
It is noted that eigenfunction, $g_n^{(0)}$, and eigenvalue, $K_n^{(0)}$, 
need to be free from $r$ by separation of variables.
In paper I, however, they depend on $\omega^2/\Omega_\bot^2$, because the boundary condition adopted
was $\omega^2/\Omega_\bot^2$-dependent. 
This means that the analytical procedures adopted in paper I had slightly internal inconsistency.
  
\begin{figure}
 \begin{center}
  \includegraphics[width=8cm]{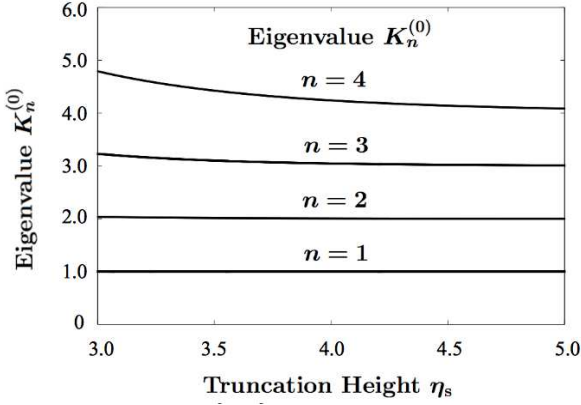} 
 \end{center}
\caption{Eigenvalues of the fundamental ($n=1$) and overtones 
($n=2,3,4...$) of g-mode oscillations as functions of $\eta_{\rm s}$. 
In the case of $n=1$, the eigenvalue is unity, independent of $\eta_{\rm s}$. 
In cases of $n\not= 1$, however, the eigenvalues are slightly larger than $n$, 
but tend to $n$ as $\eta_{\rm s}$ increases. 
}
\label{fig:sample}
\end{figure}
\begin{figure}
 \begin{center}
  \includegraphics[width=8cm]{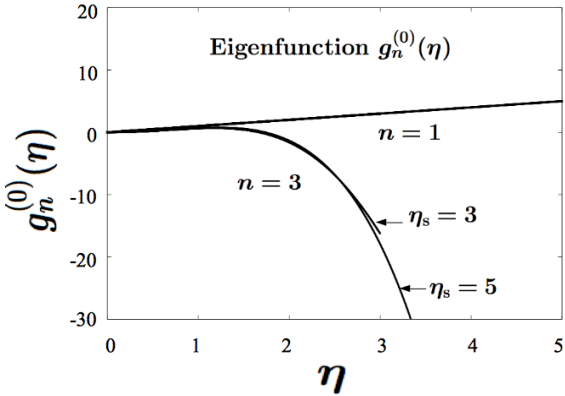} 
 \end{center}
\caption{Functional forms of eigenfunction, $g_n^{(0)}(\eta)$, of two odd modes
($n=1$, i.e.,fundamental mode, and $n=3$).
Two cases where disk boundary $\eta_{\rm s}$ is at 3.0 and 5.0 are shown.
It is noted that the eigenfunctions are terminated at $\eta=\eta_{\rm s}$,
but their functional forms are quite close to those of the corresponding Hermite
polynomials, except
in the region close to the boundary where $d^2g_n^{(0)}(\eta)/d\eta^2=0$ is imposed.   
The eigenfunction of $n=3$ in the disk with $\eta_{\rm s}=5$ is not shown till
$\eta_{\rm s}$, because it extends beyond the domain of this figure.
}
\label{fig:sample}
\end{figure}
\begin{figure}
 \begin{center}
  \includegraphics[width=8cm]{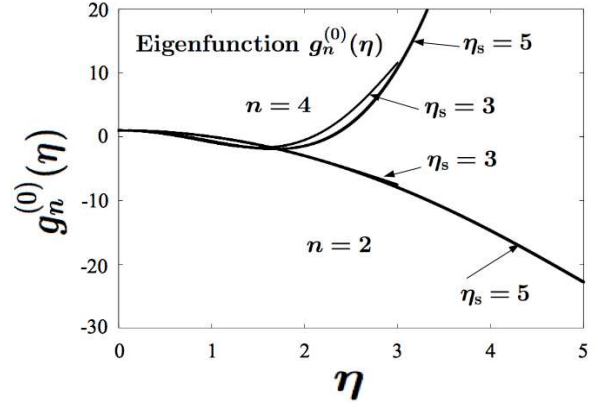} 
 \end{center}
\caption{Functional forms of eigenfunction, $g_n^{(0)}(\eta)$, of even modes, i.e.,
$n=2$ and $n=4$ modes.
Two cases where disk boundary $\eta_{\rm s}$ is at 3.0 and 5.0 are shown.
It is noted that the eigenfunctions are terminated at $\eta=\eta_{\rm s}$,
but their functional forms are quite close to the corresponding Hermite polynomials
except in the region close to
the boundary where $d^2g_n^{(0)}(\eta)/d\eta^2=0$ is imposed.  
}
\label{fig:sample}
\end{figure}

\subsection{Orthogonality of zeroth-order eigenfunctions}

In the case of infinitely extended isothermal disks, the series of eigenfunctions, say $g_n^{(0)}$, are orthogonal in the sense that 
\begin{equation}
         \int_{-\infty}^\infty{\rm exp}\biggr(-\frac{\eta^2}{2}\biggr)g_n^{(0)}g_m^{(0)} d\eta= n! (2\pi)^{1/2}\delta_{nm},
\label{orth0}
\end{equation}
because $g_n^{(0)}$ is the Hermite polynomial of order $n$.

In the case considered in paper I, the zeroth-order eigenfunctions are not rigorously 
orthogonal
(see section 3.2 in paper I), but we were reluctantly satisfied with a 
quasi-orthogonality in proceeding 
to the next order procedures in the perturbation method.
In the present case of boundary condition (\ref{BC4}), however, we have exactly an orthogonality relation of zeroth-order eigenfunctions.
This orthogonality relation is different from that in the case of infinitely extended 
isothermal disks (relation (\ref{orth0})).
The orthogonality relation in the present case is
\begin{equation}
         \int_{-\eta_{\rm s}}^{\eta_{\rm s}}{\rm exp}\biggr(-\frac{\eta^2}{2}\biggr)\frac{dg_n^{(0)}}{d\eta}
              \frac{dg^{(0)}_m}{d\eta}d\eta=0      \quad  {\rm for} \ \ n\not= m.
\label{orth1}
\end{equation}     

This orthogonality can be easily derived by taking the derivative of equation (\ref{eqg}) with respect to
$\eta$ and by changing the resulting equation in the form:
\begin{equation}
     \biggr(\frac{d^2}{d\eta^2}-\eta\frac{d}{d\eta}+K_n^{(0)}-1 \biggr)
     \frac{d g_n^{(0)}}{d\eta} =0,
\label{eqG}
\end{equation}
which is further rewritten as
\begin{equation}
    \frac{d}{d\eta}\biggr[{\rm exp}\biggr(-\frac{\eta^2}{2}\biggr)\frac{d}{d\eta}\frac{dg_n^{(0)}}{d\eta}\biggr]
           +(K_n^{(0)}-1){\rm exp}\biggr(-\frac{\eta^2}{2}\biggr)\frac{dg_n^{(0)}}{d\eta}=0.
\label{eqG2}
\end{equation}

The presence of orthogonality among the zeroth order eigenfunctions is helpful in applying the standard perturbation methods.

\section{Wave equation when $c_{{\rm A}0}^2/c_{\rm s}^2$ is taken into account}

In the case of $c_{\rm A}^2/c_{\rm s}^2\not= 0$, 
the solution of equation (\ref{eq-of-h1}) does not have such a separable form as  
$h_1(r,\eta)=g(\eta)f(r)$.
If the effects of $c_{\rm A}^2/c_{\rm s}^2\equiv c_{{\rm A}0}^2/c_{\rm s}^2\,{\rm exp}(\eta^2/2)\not= 0$ on 
oscillations are weak,
however, the terms with $c_{\rm A}^2/c_{\rm s}^2$ 
in equation (\ref{eq-of-h1}) can be treated as a small perturbation in
solving equation (\ref{eq-of-h1}).
That is, $h_1(r,\eta)$  can be approximately separated as $h_1(r,\eta)=g(\eta, r)f(r)$ with weak $r$-dependence of $g$.
This weak $r$-dependence of $g$ can be examined by a perturbation method.

The orthogonal relation (\ref{orth1}) given above shows that it is relevant to process the
perturbation method by using dependent variable, $\partial h_1/\partial\eta$, instead of $h_1$ itself. 
This is because when we proceed to examining effects of $c_{\rm A}^2/c_{\rm s}^2$ on oscillations, the perturbed part of eigenfunctions can be expanded in terms of the zeroth order eigenfunctions which are orthogonal.

To proceed to this direction, we take the derivative of equation (\ref{eq-of-h1}) with respect to $\eta$.
Neglecting the term of $(c_{\rm A}^2/c_{\rm s}^2)^2$, we have then
\begin{eqnarray}
&&    ({\cal L}-1)\frac{\partial h_1}{\partial\eta}+\frac{\omega^2H^2}{\omega^2-\kappa^2}
                      \frac{\partial^2}{\partial r^2}\biggr(\frac{\partial h_1}{\partial \eta}\biggr)
                  \nonumber     \\
&&    +\frac{c_{\rm A}^2}{c_{\rm s}^2}\frac{\Omega_\bot^2}{\omega^2-\kappa^2}
           \biggr[H^2\frac{\partial^2}{\partial r^2}\biggr({\cal L}-1+\frac{\partial^2}{\partial\eta^2}\biggr) 
           \biggr(\frac{\partial h_1}{\partial \eta}\biggr)
                 \nonumber \\
&& \hspace{30pt}         +\frac{2\omega^2-(\kappa^2-4\Omega^2)}{\omega^2}
              \frac{\partial^2}{\partial\eta^2}\biggr(({\cal L}-1)\frac{\partial h_1}{\partial\eta}\biggr)\biggr]
              \nonumber \\
&&    +\frac{c_{\rm A}^2}{c_{\rm s}^2}\frac{\Omega_\bot^2}{\omega^2-\kappa^2}
           \biggr[H^2\frac{\partial^2}{\partial r^2}\biggr(\eta{\cal L}+\eta\frac{\partial^2}{\partial\eta^2}\biggr)h_1
                 \nonumber \\
&& \hspace{30pt}         +\frac{2\omega^2-(\kappa^2-4\Omega^2)}{\omega^2}
              \eta\frac{\partial^2}{\partial\eta^2}\biggr({\cal L}h_1\biggr)\biggr].
\label{eq-of-dh1}
\end{eqnarray}

Now we write $\partial h_1(r,\eta)/\partial \eta$ as $G(\eta,r)f(r)$, i.e., 
\begin{equation}
        \frac{\partial}{\partial \eta} h_1(r,\eta)=G(\eta, r)f(r) \quad {\rm with}\quad 
                   \frac{\partial g(\eta, r)}{\partial \eta}=G(\eta, r),
\label{def-G}
\end{equation}
and divide equation (\ref{eq-of-dh1}) by $G(\eta, r)f(r)$ in order to 
approximately separate the resulting equation into two equations with a weakly $r$-dependent 
separation constant $K(r)$ as
\begin{eqnarray}
&&    \frac{1}{G}\biggr(\frac{\partial^2}{\partial \eta^2}-\eta\frac{\partial}{\partial\eta}-1\biggr)G
        +\frac{\omega^2H^2}{\omega^2-\kappa^2}\frac{1}{G}
                      \biggr(\frac{\partial^2G}{\partial r^2}+2\frac{\partial G}{\partial r}\frac{d{\rm ln}f}{dr}\biggr)
                  \nonumber     \\
&&    +\frac{c_{\rm A}^2}{c_{\rm s}^2}\frac{\Omega_\bot^2}{\omega^2-\kappa^2}\frac{1}{Gf}
           \biggr[H^2\frac{\partial^2}{\partial r^2}\biggr({\cal L}-1+\frac{\partial^2}{\partial\eta^2}\biggr)(Gf)
                 \nonumber \\
&& \hspace{30pt}         +\frac{2\omega^2-(\kappa^2-4\Omega^2)}{\omega^2}
              \frac{\partial^2}{\partial\eta^2}\biggr(({\cal L}-1)Gf\biggr)\biggr]
              \nonumber \\
&&    +\frac{c_{\rm A}^2}{c_{\rm s}^2}\frac{\Omega_\bot^2}{\omega^2-\kappa^2}\frac{1}{Gf}
           \biggr[H^2\frac{\partial^2}{\partial r^2}\biggr(\eta{\cal L}+\eta\frac{\partial^2}{\partial\eta^2}\biggr)(gf)
                 \nonumber \\
&& \hspace{30pt}         +\frac{2\omega^2-(\kappa^2-4\Omega^2)}{\omega^2}
              \eta\frac{\partial^2}{\partial\eta^2}\biggr({\cal L}(gf)\biggr)\biggr]
                 +K=0.
\label{eq-of-G}
\end{eqnarray}
and
\begin{equation}
      \frac{\omega^2}{\Omega_\bot^2}+\frac{\omega^2H^2}{\omega^2-\kappa^2}\frac{1}{f}\frac{d^2f}{dr^2}
         -K=0.
\label{eq-of-f00}
\end{equation}
It is noticed that in equation (\ref{eq-of-G}) some terms still have $g$, not $G$.

In the lowest order of approximation, i.e., ${c_{\rm A}^2}/{c_{\rm s}^2}=0$, 
we see that $G$ is a function of $\eta$ alone, and the above set of equations are reduced to
\begin{equation}
     \biggr(\frac{d^2}{d\eta^2}-\eta\frac{d}{d \eta}-1\biggr)G^{(0)}+K^{(0)}G^{(0)}=0
\label{eq-of-G0}
\end{equation}
and
\begin{equation}
      \frac{\omega^2H^2}{\omega^2-\kappa^2}\frac{d^2 f}{dr^2}
      +\biggr(\frac{\omega^2}{\Omega_\bot^2}-K^{(0)}\biggr)f=0,
\label{eqf-new}
\end{equation}
where superscript $(0)$ have been attached to $G$ and $K$ in order to emphasize that these are the zeroth order solutions.
The latter equation is the same as equation (\ref{eqf}).

Now we consider the $n$-th mode of oscillations, and write $G^{(0)}$ and $K^{(0)}$ as $G_n^{(0)}$ and $K_n^{(0)}$.
If we proceed to the next order of approximation, $G_n[=G_n^{(0)}(\eta)+G_n^{(1)}(\eta,r)]$ is 
no longer a function of $\eta$ alone.
It depends weakly on $r$.
Then, from equation (\ref{eq-of-G}) we have
\begin{equation}
   \biggr(\frac{\partial^2}{\partial \eta^2}-\eta\frac{\partial}{\partial\eta}-1+K_n^{(0)} \biggr)G_n^{(1)}= R_1 + R_2,
\label{eq-of-G1-0}
\end{equation}
where
\begin{equation}
       R_1=-\frac{\omega^2H^2}{\omega^2-\kappa^2}
          \biggr(\frac{\partial^2G_n^{(1)}}{\partial r^2}+2\frac{\partial G_n^{(1)}}
          {\partial r}\frac{d{\rm ln}f}{dr}\biggr)
\label{eq-of-G1-1}
\end{equation}
and 
\begin{eqnarray}
&&  R_2=  -\frac{c_{\rm A}^2}{c_{\rm s}^2}\frac{\Omega_\bot^2}{\omega^2-\kappa^2}
           \frac{1}{f}\biggr[H^2\frac{\partial^2}{\partial r^2}\biggr({\cal L}-1+\frac{\partial^2}{\partial\eta^2}\biggr) 
                   (G_n^{(0)}f)
                 \nonumber \\
 &&\hspace{30pt}         +\frac{2\omega^2-(\kappa^2-4\Omega^2)}{\omega^2}
              \frac{\partial^2}{\partial\eta^2}\biggr(({\cal L}-1)G_n^{(0)}f\biggr)\biggr]
              \nonumber \\
&&    -\frac{c_{\rm A}^2}{c_{\rm s}^2}\frac{\Omega_\bot^2}{\omega^2-\kappa^2}
           \frac{1}{f}\biggr[H^2\frac{\partial^2}{\partial r^2}\biggr(\eta{\cal L}
           +\eta\frac{\partial^2}{\partial\eta^2}\biggr)(g_n^{(0)}f)
                 \nonumber \\
&& \hspace{30pt}         +\frac{2\omega^2-(\kappa^2-4\Omega^2)}{\omega^2}
              \eta\frac{\partial^2}{\partial\eta^2}\biggr({\cal L}(g_n^{(0)}f)\biggr)\biggr]
                   \nonumber \\
&&              -K_n^{(1)}G_n^{(0)},
\label{eq-of-G1-2}
\end{eqnarray}
and from equation (\ref{eq-of-f00}) 
we have 
\begin{equation}
      \frac{\omega^2}{\Omega_\bot^2}f+\frac{\omega^2H^2}{\omega^2-\kappa^2}\frac{d^2f}{dr^2}
         -(K_n^{(0)}+K_n^{(1)})f=0.
\label{eq-of-f}
\end{equation}

To solve equation (\ref{eq-of-G1-0}) by the standard perturbation method, $G_n^{(1)}$ is 
now expanded in terms of the set of zeroth order orthogonal eigenfunctions,  
say $G^{(0)}_m$, as 
\begin{equation}  
             G_n^{(1)}(\eta, r)=\sum_{m}a_{m}(r)G_m^{(0)}(\eta),
\label{expansion_G1}
\end{equation}
where $a_m(r)$'s are expansion coefficients, and
$m$'s needed here are odd integers alone when odd modes ($g_n$ is an odd function of 
$\eta$, but $G_n$ is an even function of $\eta$) are considered, while they are even 
integers alone when even modes are considered.
It is noted that each of $G_m^{(1)}$'s given above satisfies boundary condition (\ref{BC4}) and is orthogonal each other (equation (\ref{orth1})).

Substitution of equation (\ref{expansion_G1}) into equation (\ref{eq-of-G1-0})
shows that the term of $a_nG_n^{(0)}$ vanishes on the lefthand side of equation
(\ref{eq-of-G1-0}), because $G_n^{(0)}$ is the eigenfunction of 
equation (${\partial}/\partial\eta^2-\eta\partial/\partial\eta-1+K_n^{(0)})G=0$.
This means that the righthand side of (\ref{eq-of-G1-0}) needs to be orthogonal
to $G_n^{(0)}$ (i.e., solvability condition).
This solvability condition is given in the form:
\begin{equation}
      \int_{-\eta_{\rm s}}^{\eta_{\rm s}}{\rm exp}\biggr(-\frac{\eta^2}{2}\biggr) 
             G_n^{(0)}(\eta)\cdot (R_1+R_2)d\eta=0.
\label{solvability0}
\end{equation} 

In the terms of $R_1$ given by equation (\ref{eq-of-G1-1}), $a_nG_n^{(0)}$ is involved, since $G_n^{(1)}$ is expanded
in terms of the zeroth order eigenfunctions. 
The terms with $a_nG_n^{(0)}$, however, can be taken to be zero in consideration of solvability condition (\ref{solvability0}), 
since such terms can be regarded to be already included in the zeroth order 
eigenfunction $G_n^{(0)}$ (i.e., normalization condition).
The other terms of $a_mG_m^{(0)}$ ($m\not= n$) in $G_n^{(1)}$ do not contribute in 
expression for the solvability condition (\ref{solvability0}), 
since $G_m^{(0)}$'s of $m \not= n$ are orthogonal to $G_n^{(0)}$.    
Hence, the solvability condition (\ref{solvability0}) is reduced to 
\begin{equation}
      \int_{-\eta_{\rm s}}^{\eta_{\rm s}}{\rm exp}\biggr(-\frac{\eta^2}{2}\biggr) 
             G_n^{(0)}(\eta)\cdot R_2d\eta=0.
\label{solvability1}
\end{equation}

Before writing down solvability condition (\ref{solvability1}) explicitly, we notice 
that the some terms on the 
righthand side of equation (\ref{eq-of-G1-2}) can be simplified.
That is, we have
\begin{equation}
     ({\cal L}-1)G_n^{(0)}=\biggr(\frac{\omega^2}{\Omega_\bot^2}-K_n^{(0)}\biggr)G_n^{(0)},
\end{equation}
and
\begin{equation}
     {\cal L}g_n^{(0)}=\biggr(\frac{\omega^2}{\Omega_\bot^2}-K_n^{(0)}\biggr)g_n^{(0)}.
\end{equation}
Then, $R_2$ given by equation (\ref{eq-of-G1-2}) can be reduced to
\begin{eqnarray}
&&  R_2=  -\frac{c_{\rm A}^2}{c_{\rm s}^2}\frac{\Omega_\bot^2}{\omega^2-\kappa^2}
           \biggr[H^2\frac{d^2f}{fdr^2}
           \biggr(\frac{d^2}{d\eta^2}+
                 \frac{\omega^2-K_n^{(0)}\Omega_\bot^2}{\Omega_\bot^2}\biggr)G_n^{(0)}
                 \nonumber \\
&&\hspace{30pt}         +\frac{2\omega^2-(\kappa^2-4\Omega^2)}{\omega^2}
           \frac{\omega^2-K_n^{(0)}\Omega_\bot^2}{\Omega_\bot^2}
             \frac{d^2G_n^{(0)}}{d\eta^2}
              \nonumber \\
&&    -\frac{c_{\rm A}^2}{c_{\rm s}^2}\frac{\Omega_\bot^2}{\omega^2-\kappa^2}
           \biggr[H^2\frac{d^2f}{fdr^2}\eta
            \biggr(\frac{d^2}{d\eta^2}
                 +\frac{\omega^2-K_n^{(0)}\Omega_\bot^2}{\Omega_\bot^2}\biggr) g_n^{(0)}
                 \biggr)
                 \nonumber \\
&& \hspace{30pt}         +\frac{2\omega^2-(\kappa^2-4\Omega^2)}{\omega^2}
              \frac{\omega^2-K_n^{(0)}\Omega_\bot^2}{\Omega_\bot^2}
              \eta\frac{d^2g_n^{(0)}}{d\eta^2}
                   \nonumber \\
&&              -K_n^{(1)}G_n^{(0)},
\label{eq-of-G1-3}
\end{eqnarray}

The righthand side of equation ({\ref{eq-of-G1-3}) is further simplified,
since $d^2f/fdr^2$ can be expressed in terms of the zeroth order one:
\begin{equation}
     \frac{\omega^2H^2}{\omega^2-\kappa^2}\frac{1}{f}\frac{d^2f}{dr^2}
        +\biggr(\frac{\omega^2}{\Omega_\bot^2}-K_n^{(0)}\biggr)=0.
\label{eq-of-f0}
\end{equation}
Then, $R_2$ is finally reduced to
\begin{eqnarray}
&&   R_2=  -\frac{c_{\rm A}^2}{c_{\rm s}^2}\frac{\omega^2-K_n^{(0)}\Omega_\bot^2}          
               {\omega^2-\kappa^2}
               \nonumber  \\
&&         \times\biggr[-\frac{\omega^2-\kappa^2}{\omega^2}\frac{\omega^2-K_n^{(0)}
               \Omega_\bot^2}{\Omega_\bot^2}(G_n^{(0)}+\eta g_n^{(0)})
               \nonumber  \\
&&\hspace{15pt}            +\frac{\omega^2+4\Omega^2}{\omega^2}
      \biggr(\frac{d^2G_n^{(0)}}{d\eta^2}+\eta\frac{d^2g_n^{(0)}}{d\eta^2}\biggr)\biggr]
            -K_n^{(1)}G_n^{(0)}.
\label{R2-1}
\end{eqnarray}

Using the above expression for $R_2$, we can write down solvability condition (\ref{solvability1}) as
\begin{eqnarray}
&&      K_n^{(1)}\biggr\langle{\rm exp}\biggr(-\frac{\eta^2}{2}\biggr)G_n^{(0)}G_n^{(0)}
                \biggr\rangle
            \nonumber   \\
&&        =  \frac{c_{{\rm A}0}^2}{c_{\rm s}^2}\frac{\omega^2-K_n^{(0)}\Omega_\bot^2}
               {\omega^2}
               \nonumber  \\
&&        \times \biggr[\frac{\omega^2-K_n^{(0)}\Omega_\bot^2}{\Omega_\bot^2}
                   \biggr(\biggr\langle G_n^{(0)}G_n^{(0)}\biggr\rangle
                         +\biggr\langle \eta G_n^{(0)}g_n^{(0)}\biggr\rangle\biggr)
             \nonumber  \\
&&                  -\frac{\omega^2+4\Omega^2}{\omega^2-\kappa^2}
            \biggr(\biggr\langle G^{(0)}\frac{d^2G_n^{(0)}}{d\eta^2}\biggr\rangle
            +\biggr\langle\eta G_n^{(0)}\frac{dG_n^{(0)}}{d\eta}\biggr\rangle\biggr)
       \biggr],
\label{K1}
\end{eqnarray} 
where $\langle \ \ \rangle$ represents the integration of the quantities inside 
$\langle \ \ \rangle$ in terms of $\eta$ from $-\eta_{\rm s}$ to $\eta_{\rm s}$, e.g.,
\begin{equation}
     \langle G_n^{(0)}G_n^{(0)}\rangle=\int_{-\eta_{\rm s}}^{\eta_{\rm s}}
               G_n^{(0)}G_n^{(0)}d\eta.
\end{equation}

By using $K_n^{(1)}$ given by equation (\ref{K1}), 
we can write the wave equation 
(\ref{eq-of-f}) finally in the form:
\begin{equation}
   H^2 \frac{d^2f}{dr^2}+k^2f=0,
\label{final-wave-eq}
\end{equation}
where
\begin{equation}
     k^2=\frac{(\omega^2-\kappa^2)(\omega^2-K_n^{(0)}\Omega_\bot^2)}
          {\omega^2 \Omega_\bot^2}(1+Q),
\label{final-of-f2}
\end{equation}
and
\begin{equation}
       Q(r;\omega)=
      -\frac{c_{{\rm A}0}^2}{c_{\rm s}^2}\frac{\Omega_\bot^2}{\omega^2}
           \biggr(A\frac{\omega^2-K_n^{(0)}\Omega_\bot^2}{\Omega_\bot^2}
        -B\frac{\omega^2+4\Omega^2}{\omega^2-\kappa^2}\biggr). 
\label{final-of-f3}
\end{equation} 
Here, $A$ and $B$ are given, respectively, by
\begin{equation}
    A=\frac{\langle G_n^{(0)}G_n^{(0)}\rangle
               +\langle\eta G_n^{(0)}g_n^{(0)}\rangle}
       {\langle{\rm exp}({-\eta^2/2}) G_n^{(0)}G_n^{(0)}\rangle}.
\label{def-A}
\end{equation}
and
\begin{equation}
     B=\frac{\langle G_n^{(0)}d^2G_n^{(0)}/d\eta^2\rangle
               +\langle\eta G_n^{(0)}dG_n^{(0)}/d\eta\rangle}
       {\langle{\rm exp}({-\eta^2/2}) G_n^{(0)}G_n^{(0)}\rangle}.
\label{def-B} 
\end{equation}      
The final wave equations, (\ref{final-wave-eq}) -- (\ref{final-of-f3}), should be compared with
the wave equation derived with a different boundary condition in paper I.
It is noticed that both of them have quite similar forms, although
detailed expressions for $A$ and $B$ are slightly different.\footnote{
Hereafter, $B$ is the quantity defined by equation (\ref{def-B}), and not
the strength of magnetic fields.
}

\begin{table}
	\centering
	\caption{Values of $A$, $B$, and the conditions of $Ac_{{\rm A}0}^2/c_{\rm s}^2<1$.}
	\label{tab:example_table}	
	\begin{tabular}{lcrrc} 
	\hline
   $\eta_{\rm s}$ & oscillation   &  $A$ & $B$& condition of     \\
                  &  mode         &        & & $Ac_{{\rm A}0}^2/c_{\rm s}^2<1$ \\
   \hline   
   3.0  &  $n=1$   & 19.30 &  0   &  $c_{{\rm A}0}^2/c_{\rm s}^2< 5.2\times 10^{-2}$  \\
	         &  $n=2$   & 21.32 &  0.323  & $c_{{\rm A}0}^2/c_{\rm s}^2< 4.7\times 10^{-2}$ \\
	         &  $n=3$   & 46.96 & 22.52     & $c_{{\rm A}0}^2/c_{\rm s}^2< 2.1\times 10^{-2}$  \\
	         &  $n=4$   & 32.82 & 9.217   & $c_{{\rm A}0}^2/c_{\rm s}^2< 3.0\times 10^{-2} $ \\
   \hline	
  	    5.0  &  $n=1$   & $1.19\times 10^2$ & 0  &       $c_{{\rm A}0}^2/c_{\rm s}^2< 8.4\times 10^{-3}$ \\
	         &  $n=2$   & $2.09\times 10^2$ &$-5.86\times 10^1$ &     $c_{{\rm A}0}^2/c_{\rm s}^2< 4.8\times 10^{-3}$ \\
             &  $n=3$   & $2.93\times 10^3$ &$7.20\times 10^2$ &     $c_{{\rm A}0}^2/c_{\rm s}^2< 3.4\times 10^{-4}$ \\
             &  $n=4$   & $3.31\times 10^3$ &$-1.56\times 10^2$ &   $c_{{\rm A}0}^2/c_{\rm s}^2< 3.0\times 10^{-4}$ \\
 \hline	    

	\end{tabular}
\end{table}

\section{Self-trapping of g-mode Oscillations}

The set of equations (\ref{final-wave-eq}) -- (\ref{final-of-f3}) finally represents
the wave motions for axisymmetric g-mode oscillations in the case where the
effects of vertical magnetic fields are taken into account as perturbations.
Since the effects of $c_{{\rm A}0}^2/c_{\rm s}^2$ have been taken into account as
perturbations, $Q$ defined by equation (\ref{final-of-f3}) is necessary to be 
smaller than unity in magnitude.
Roughly speaking, this condition is written as $Ac_{{\rm A}0}^2/c_{\rm s}^2<1$.
The condition of $Ac_{{\rm A}0}^2/c_{\rm s}^2<1$ is presented in the last column 
of table 1.
As is shown in the table, $c_{{\rm A}0}^2/c_{\rm s}^2$ needs to be smaller than 
a certain value.
The value is smaller as $n$ and $\eta_{\rm s}$ become larger.
The reason of this trend is obvious, since in the higher overtone (i.e, $n$ is large) 
and in disks with large $\eta_{\rm s}$, the main oscillation region shifts 
to the surface zone of disks (see figures 2 and 3). 

Our purpose here is to examine how the propagation domain of g-mode oscillations
on the frequency-radius diagram (propagation diagram) is modified by
the presence of $Q$ defined by equation (\ref{final-of-f3}).
This examination should be made in the framework that 
the absolute value of $Q$ is smaller than
(or at most equal to) unity, since $Q$ is a quantity obtained by a perturbation method. 

In frequency-radius diagram the propagation domain of oscillations with frequency 
$\omega$ (i.e., propagation domain in propagation diagram) is the domain where 
$k^2(r; \omega)>0$.
Equation (\ref{final-of-f2}) confirms that if there is no magnetic field, 
the propagation domain is given 
by the area below the curve of $\kappa(r)$, as is well known.\footnote{
Notice that $\omega^2-K_n^{(0)}\Omega_\bot^2 <0$ when $\omega^2<\kappa^2$.
}
In the frequency-radius diagram the curve of $\kappa(r)$ has a maximum, say
$\kappa_{\rm max}$, at a certain radius, say $r_{\rm max}$, and $\kappa(r)$ 
decreases both inward and outward.
Inside $r_{\rm max}$, $\kappa$ decreases to vanish at the disk inner edge.
Since the curve of $\kappa(r)$ has the maximum, the oscillations whose frequency is
lower than $\kappa_{\rm max}$ are trapped in a certain radius range whose 
inner and outer boundaries are specified by the radii of $\omega=\kappa$.
This is the self-trapping of g-mode oscillations in non-magnetized disks.

Let us now consider the effects of the term with $A$ in $Q$ on $k^2$. 
This term is positive in the domain of $\omega^2<\kappa^2$ if $A>0$.
Since $A$ is always positive in all cases considered,
we see that the term with $A$ in equation (\ref{final-of-f3}) is always positive
and work in the direction to make $k^2$ larger than that in the case of 
$c_{\rm A}^2/c_{\rm s}^2=0$.
This means that the term of $A$ does not work so as to change the boundary of the
propagation region from $\omega^2=\kappa^2$. 

Next, let us consider the effects of $B$ on $k^2$.
Before discussing generally the effects, 
it is important to notice
that in the case of the fundamental g-mode oscillations (i.e., $n=1$),
we have $B=0$, independent of disk parameters (see table 1).
This means that the trapping domain of the oscillations is unchanged from that of 
non-magnetized infinitely extended isothermal disks.
$B=0$ comes from the fact that the eigenfunction $g_1^{(0)}(\eta)$ is given by
$g_1^{(0)}(\eta)=\eta$, and thus $G_1^{(0)}(\eta)=1$ and $dG_1^{(0)}/d\eta=0$, leading to 
$B=0$, independent of $\eta_{\rm s}$.

In the cases of other oscillation modes situations are changed.
First we consider cases of $B>0$.\footnote{As shown in table 1, $B$ is positive 
except in the cases of even modes in disks with
large $\eta_{\rm s}$, say $\eta_{\rm s}=5$.
}
In this case the term with $B$ in $Q$ is negative when $\omega^2<\kappa^2$.
This means that the term with $B$ in $Q$ acts in the direction to make $k^2$ smaller.
In other words, the term with $B$ acts in the direction to shrink the propagation 
domain, compared with in the case where the boundary is specified by 
the curve of $\kappa(r)$.
Some typical examples of the propagation domain in cases of $B>0$ are shown 
in figures 4 to 7 for some oscillation modes ($n$) in disks with
various $c_{{\rm A}0}^2/c_{\rm s}^2$, $\eta_{\rm s}$ and spin parameter $a_*$ of
the central source.
 
Figures 4 and 5 are for disks with $c_{{\rm A}0}^2/c_{\rm s}^2=0.001$ and
$c_{{\rm A}0}^2/c_{\rm s}^2=0.01$, respectively, for four oscillations modes
($n=$1, 2, 3, and 4).
The truncation height of disks has been taken to be $\eta_{\rm s}=3.0$.
In figure 6 the truncation height, $\eta_{\rm s}$, is taken to be $\eta_{\rm s}=5.0$
and the effects of $c_{{\rm A}0}^2/c_{\rm s}^2$ on propagation region
have been examined for oscillation mode of $n=3$.
Effects of spin of central sources are examined in figure 7 for 
the oscillation mode of $n=3$ in disks with $c_{{\rm A}0}^2/c_{\rm s}^2=0.01$.

In the case of $B<0$ the situation is opposite, and the propagation domain expands
on the propagation diagram.
Due to this, the propagation domain can spread inwards beyond the barrier of epicyclic
frequency (the curve of $\kappa(r)$), i.e., the self-trapping of g-mode oscillations is 
destroyed in the inner region of disks.
Such examples are in figure 8, where even modes of $n=2$ and $n=4$ are considered
in the disks with $\eta_{\rm s}=5.0$ and $c_{{\rm A}0}^2/c_{\rm s}^2 =10^{-3}$.
In the case of oscillations of $n=4$, the propagation region penetrates inwards beyond
the barrier of $\kappa(r)$ for oscillations whose frequency is relatively low.
That is, the self-trapping is destroyed for low frequency oscillations.
In the case of oscillations of $n=2$, the self-trapping is completely destroyed
for whole frequencies of g-mode oscillations.
It is noted, however, that the cases of $B<0$
seem to appear only for even modes of oscillations in disks with large $\eta_{\rm s}$.   
  
\begin{figure}
 \begin{center}
  \includegraphics[width=8cm]{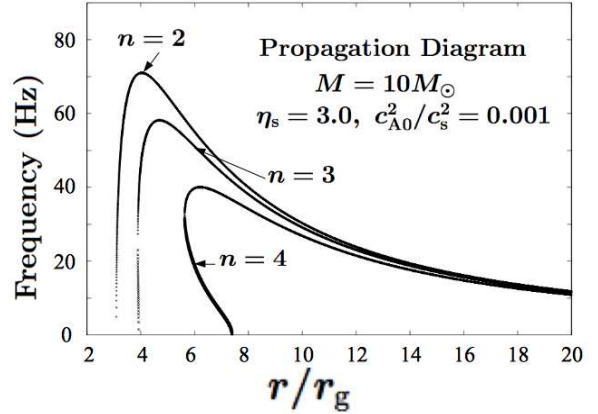} 
 \end{center}
\caption{Propagation diagram showing the propagation domain of axisymmetric g-mode 
oscillations in cases where the central source is a ten-solar-mass object
($10M_\odot$) with no spin ($a_*=0$) and the disk is subject to uniform vertical 
magnetic fields of $c_{{\rm A}0}^2/c_{\rm s}^2=0.001$.
The oscillation modes considered are $n=2$, $n=3$, and $n=4$.
The boundary curve of the propagation domain of $n=2$ oscillations is close to 
the curve of $\kappa(r)$.
Hence, the curve for $n=1$, which is identical to $\kappa(r)$, is not shown here.
The region below each curve is the propagation domain for each mode.  
It is shown that the domain shrinks with increase of $n$.
This is because contributions of surface region of disks on oscillations increase 
with increase of mode number of oscillations.
}
\label{fig:sample}
\end{figure}
\begin{figure}
 \begin{center}
  \includegraphics[width=8cm]{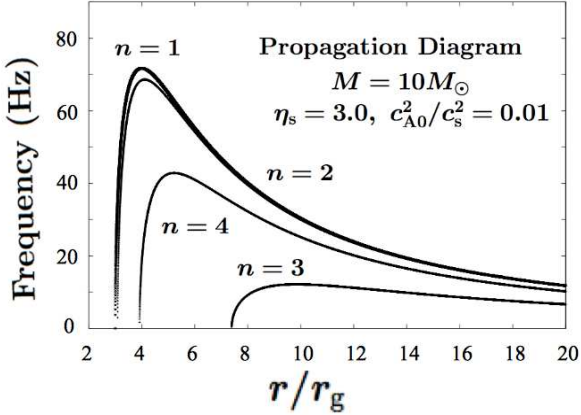} 
 \end{center}
\caption{The same as figure 4, except for $c_{{\rm A}0}^2/c_{\rm s}^2=0.01$.
In this figure the curve of $n=1$ (the curve of $\kappa(r)$) is also shown, 
because two curves of $n=1$ and $n=2$ are separated a little on the diagram
due to a large value of $c_{{\rm A}0}^2/c_{\rm s}^2$. 
It is noticed that the order of curves of $n=2$, $n=3$, and $n=4$ on the plane
is not monotonic.   
}
\label{fig:sample}
\end{figure}
\begin{figure}
 \begin{center}
  \includegraphics[width=8cm]{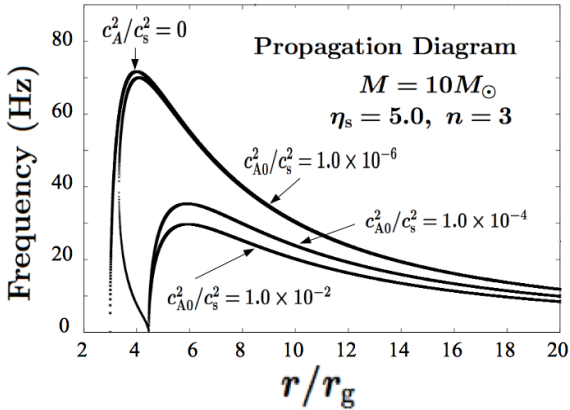} 
 \end{center}
\caption{Propagation diagram showing propagation domains of g-mode oscillations
of $n=3$ in the case where the transition height to corona is at $\eta_s=5.0$.
The boundary curves of the propagation domain are shown for three cases 
where the disk has vertical magnetic 
fields of $c_{{\rm A}0}^2/c_{\rm s}^2=1.0\times 10^{-6}$, $1.0\times 10^{-4}$,
and $1.0\times 10^{-2}$.
The curve for $c_{{\rm A}0}^2/c_{\rm s}^2=0$ is also presented for comparison.  
The value of $c_{{\rm A}0}^2/c_{\rm s}^2=1.0\times 10^{-2}$ is too large beyond 
the limit of applicability of the perturbation method (see table 1).
For comparison, however, the curve in this case is also shown in this figure.
Approaching of curves to a common point on the horizontal axis is due to the fact that
in the cases of too large value of $c_{{\rm A}0}^2/c_{\rm s}^2$, the terms with $A$ and
$B$ in the large brackets of equation (\ref{final-of-f2}) becomes larger than unity
and the curves depend little on $c_{{\rm A}0}^2/c_{\rm s}^2$.
}
\label{fig:sample}
\end{figure}
\begin{figure}
 \begin{center}
  \includegraphics[width=8cm]{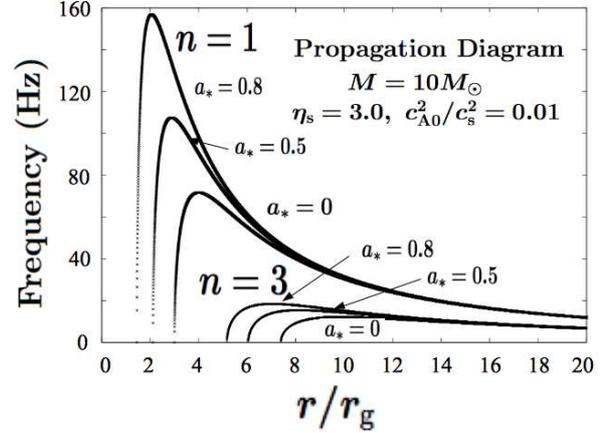} 
 \end{center}
\caption{Examples showing how the propagation domain is changed by spin of
the central source.
Two oscillation modes of $n=1$ and $n=3$ are shown for three
cases of $a_*=0$, 0.5, and 0.8, respectively.
In the former case of $n=1$, the boundary curves are nothing but the curves of 
$\kappa(r)$. 
}
\label{fig:sample}
\end{figure}
\begin{figure}
 \begin{center}
  \includegraphics[width=8cm]{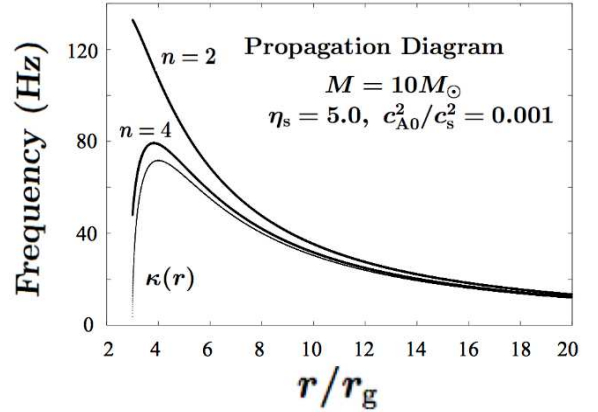} 
 \end{center}
\caption{Examples showing destruction of self-trapping domain of g-mode oscillations.
Even oscillation modes ($n=2$ and $n=4$) are shown in disks where the transition 
height is as high as $\eta_{\rm s}=5.0$.
For comparison the curve of $\kappa(r)$ is shown with a thin curve. 
}
\label{fig:sample}
\end{figure}

\section{Discussion}

In geometrically thin, non-magnetized relativistic disks the g-mode oscillations 
are self-trapped in the inner region of disks (Okazaki et al. 1987).
Fu and Lai (2005), however, emphasized that if the disks are subject to poloidal 
magnetic fields, this self-trapping of g-mode oscillations is destroyed even if
the magnetic fields are weak.
Their analyses are, however, based on an approximate treatment of $z$-dependence of oscillations.
In geometrically thin disks the Alfv\'{e}n speed increases in the vertical direction
unless strength of magnetic fields decreases sharply in the vertical direction, since
the gas density decreases strongly in the vertical direction.
In other words, the disk has a strong inhomogeneous structure in the vertical direction
(i.e., increase of Alfv\'{e}n speed and decrease of gas density in the vertical direction).
In such disks, a careful analysis is necessary to treat oscillations with node(s) 
in the vertical direction (g-mode oscillations belong to such oscillations).

In previous papers (paper I and its correction by Kato 2017b) 
we have considered a disk-corona system and 
imposed a boundary condition between the disk and the corona, in order to avoid  
mathematical difficulties in treating oscillations in strongly inhomogeneous systems.
This procedure was, however, not only for simplicity, but also to reply to observational evidences (Remillard 2005) that high-frequency QPOs in black-hole binary systems
are observed in the phase
where the gases consist both of cold and hot parts.
Remillard (2005) further mentions that frequencies of high frequency QPOs 
seem to be determined by cold disks, but the luminosity variations
seem to be associated with high temperature regions.

In this paper we considered again a disk-corona system, but adopted a boundary 
condition different from that in the previous papers.
This is because the boundary condition adopted in paper I 
seems to be not always 
proper to treat the present disk-corona transition problem, as discussed in 
section 2.3.
In addition, the boundary condition adopted in this paper allowed us to
use the standard perturbation method without introducing any additional approximate procedure, although this does not mean physical relevance of present boundary condition.
That is, the zeroth order eigenfunctions satisfying the boundary condition are a set of 
orthogonal functions and thus the next-order perturbed quantities
can be expanded by the set of the zeroth order eigenfunctions.

An interesting result obtained by this paper is
that the fundamental g-mode mode ($n=1$) oscillation is always self-trapped and
the trapped domain on the propagation diagram (i.e., frequency-radius diagram)
is the same as that in the case of non-magnetized disks.
This characteristics is free from truncation height $\eta_{\rm s}$, and also from
strength of magnetic fields.
Higher overtones ($n=$2, 3,...) are also self-trapped except for some exceptional cases
of even overtones ($n=$2, 4, ...) in disks with large $\eta_{\rm s}$ (see figure 8).
The trapped regions are, however, sensitive to strength of magnetic fields 
($c_{\rm A}^2/c_{\rm s}^2$) and disk thickness ($\eta_{\rm s}$) as well as oscillation
mode ($n$), as shown in figures 4 to 7.

There are some issues concerning the validity of the results
of this paper.
In figures 4 - 8 we have presented numerical results showing propagation domains.
In these calculations the boundary of the propagation domain is actually the boundary
of $1-Q=0$.
It is uncertain, however, whether the results obtained by the perturbation method is
accurately applicable till the case where $Q$ is as large as $Q=1$.
Next, in some cases we have obtained bounded propagation domains 
in low frequency region (e.g., figure 7).
This is of interest in relation to observations.
In the case where frequency of oscillations is rather low, however, the boundary
condition, $\delta p=0$ may be better than the condition adopted in this paper (see arguments in section 2.3).
If so, the presence of possible low frequency trapped mode may be false.

Finally, it should be noticed that the results obtained by the 
present boundary condition are rather different from
those obtained in paper I by using the boundary condition of $\delta p=0$.
We suppose that the boundary condition adopted in paper I (i.e., $\delta p=0$) will be
less proper than that adopted in this paper.
However, it is unclear how much the present boundary condition prefers to
the boundary condition adopted in paper I.
Considering that the final results of trapping are rather sensitive to boundary
conditions adopted, further careful examinations are expected in future.
A honest and tactless way is to derive wave solutions both in disk and corona and to
fit these wave solutions at the transition layer.

We think self-trapped g-mode oscillations will be excited on disks by
stochastic effects of turbulence, as mentioned in paper I. 

The author thanks Dr. Jiri Hor\'{a}k for invaluable comments 
on boundary conditions as the referee, by which ambiguous arguments 
on boundary conditions in the original manuscript have been improved.

\end{document}